\documentclass{article}
\usepackage[T1]{fontenc} 
\usepackage[utf8]{inputenc} 
\usepackage{ismir,amsmath,cite,url}
\usepackage{graphicx}
\usepackage{microtype}
\usepackage{booktabs}
\usepackage{url}

\usepackage{color}

\usepackage[firstpage]{draftwatermark}
\definecolor{lightgray}{rgb}{0.9,0.9,0.9}
\definecolor{darkgray}{rgb}{0.4,0.4,0.4}
\SetWatermarkFontSize{12pt}
\SetWatermarkScale{1.1}
\SetWatermarkAngle{90}
\SetWatermarkHorCenter{202mm}
\SetWatermarkVerCenter{170mm}
\SetWatermarkColor{darkgray}
\SetWatermarkText{Late-Breaking / Demo Session Extended Abstract, ISMIR 2024 Conference}

\def\lbd{}	

\title{How does the teacher rate? Observations from the NeuroPiano dataset}

\multauthor
{Huan Zhang$^1$ \hspace{0.5cm} Vincent Cheung$^2$ \hspace{0.5cm} Hayato Nishioka$^2$  \hspace{0.5cm} Simon Dixon$^1$ \hspace{0.5cm} Shinichi Furuya$^2$} { 
 $^1$ Queen Mary University of London, Centre for Digital Music\\
$^2$Sony Computer Science Laboratories, Tokyo, Japan
}

\def\authorname{H. Zhang, V.K.M. Cheung, H. Nishioka, S. Dixon, S. Furuya}

\begin{document}

\maketitle

\begin{abstract}
This paper provides a detailed analysis of the NeuroPiano dataset, which comprise 104 audio recordings of student piano performances accompanied with 2255 textual feedback and ratings given by professional pianists. We offer a statistical overview of the dataset, focusing on the standardization of annotations and inter-annotator agreement across 12 evaluative questions concerning performance quality. We also explore the predictive relationship between audio features and teacher ratings via machine learning, as well as annotations provided for text analysis of the responses.
\end{abstract}
\vspace{-10pt}
\section{Introduction} \label{sec:introduction}

Music Information Retrieval (MIR) has become instrumental in enhancing music education by enabling personalized learning experiences and automating feedback mechanisms\cite{Zhang2024FromPiano, Kim2022OverviewContext, Eremenko2020PerformanceLearning, Morsi2023SoundsPerformances}. Meanwhile, accurately imitating human teachers' feedback \cite{Zhang2024LLaQoAssessment,Morsi2024SimulatingLearning, Matsubara2021CROCUSCritiques} has been a central goal in MIR-assisted music education, particularly in the context of instrumental performance that involves expressive nuances \cite{Zhang2024DExterModels, Lerch2020AnAnalysis, Zhang2023DisentanglingPerformance, Cancino-Chacon2018ComputationalReview, Zhang2022ATEPPPerformance}. Moreover, studies have demonstrated the feasibility of using deep-learning based models to assess performance quality objectively and consistently \cite{Huang2020Score-informedAssessment, Pati2018AssessmentNetworks, Zhang2021LearnAssessment, Jin2023OrderPerformance, Parmar2021PianoAssessment}. This paper examines the NeuroPiano dataset, a collection of student piano performances of technical exercises, annotated with feedback in audio, textual, and rating score modalities.

In this report, we give a statistical overview of the dataset's content regarding audio, text, and score modalities, and explore their relationships. Starting with a detailed examination of rating consistency \cite{Jiang2023ExpertFeedback} and distribution, we also annotate the key concepts that manifested in the textual feedback. We also attempt to predict the teacher's rating from audio content. By analyzing how different modalities correlate with teacher assessments, this study contributes to the ongoing discussion about the effectiveness of MIR technologies in educational contexts. 



The NeuroPiano dataset\footnote{https://huggingface.co/datasets/anusfoil/NeuroPiano-data}, recorded and annotated by the Music Excellence Project at Sony CSL, Tokyo\footnote{https://www.sonycsl.co.jp/tokyo/10996/}, comprise 104 on-site audio recordings from 39 advanced student pianists performing six standardized technical exercises (including scales, arpeggios, dyads, block chords, octaves) on a Shigeru Kawai grand piano. Each recorded performance is associated with 12 questions addressing multiple performance dimensions from tempo to dynamics to articulation, and annotated by 45 professional pianists. Annotators answered each question by providing a textual response in Japanese, as well as a rating on a 6-point scale.

\vspace{-5pt}

\begin{figure}[h]
    \centering
    \includegraphics[width=\linewidth]{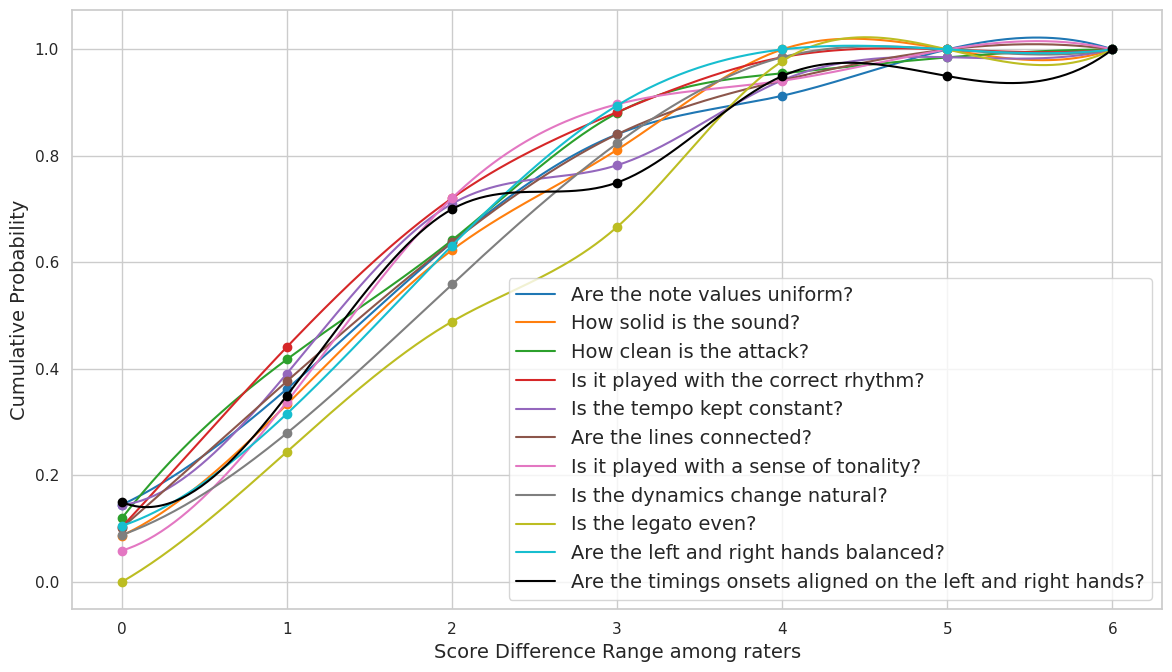}
    \includegraphics[width=\linewidth]{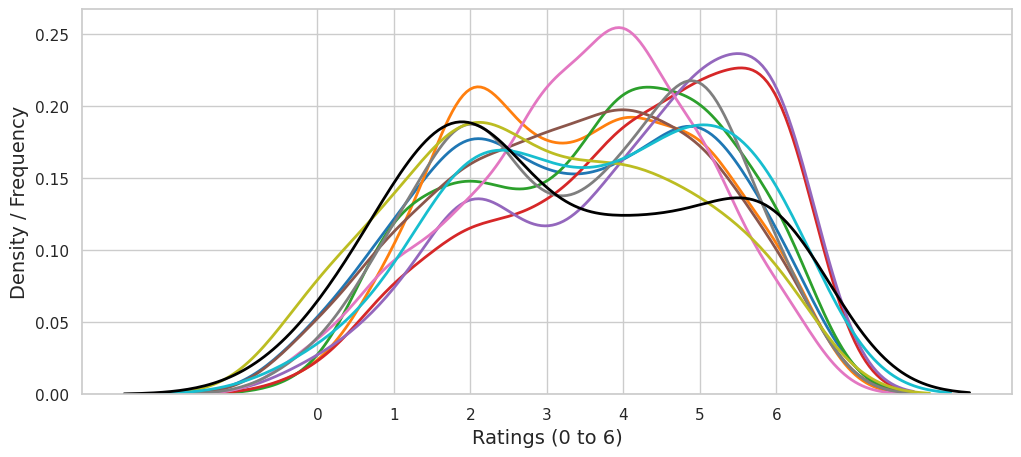}
    \caption{Top: Cumulative distribution of the raters' score difference by each question (smoothed); Bottom: Rating distribution (KDE) by each question (sharing same legend 
    )}
    \label{fig:dist}
\end{figure}

\begin{table*}[t]
\centering
\scriptsize
\begin{tabular}{@{}clcccccc@{}}  
\toprule
&  &  & \multicolumn{2}{c}{Symbolic feats.} & \multicolumn{2}{c}{Audio feats.} &  \\
\cmidrule(lr){4-5} \cmidrule(lr){6-7}
idx & Question & Support & MAE & model & MAE & model & Dummy-MAE \\
\midrule
1 & Are the note values uniform? & 91 & \textbf{1.13} & \texttt{RF(n=100)} & 1.27 & \texttt{RF(n=200)} & 1.31 \\
2 & How solid is the sound? & 90 & 1.14 & \texttt{SVR(C=0.1)} & \textbf{1.14} & \texttt{RF(n=200)} & 1.16 \\
3 & How clean is the attack? & 95 & \textbf{1.06} & \texttt{SVR(C=0.1)} & 1.11 & \texttt{SVR(C=0.1)} & 1.21 \\
4 & Is it played with the correct rhythm? & 94 & \textbf{1.16} & \texttt{SVR(C=1)} & 1.21 & \texttt{SVR(C=0.1)} & 1.27 \\
5 & Is the tempo kept constant? & 88 & \textbf{1.23} & \texttt{SVR(C=0.1)} & 1.25 & \texttt{SVR(C=0.1)} & 1.31 \\
6 & Are the lines connected? & 90 & \textbf{1.15} & \texttt{SVR(C=1)} & 1.26 & \texttt{SVR(C=0.1)} & 1.26 \\
7 & Is it played with a sense of tonality? & 95 & 1.04 & \texttt{SVR(C=0.1)} & \textbf{1.02} & \texttt{SVR(C=0.1)} & 1.05 \\
8 & Is the dynamics change natural? & 89 & \textbf{1.27} & \texttt{SVR(C=0.1)} & 1.30 & \texttt{SVR(C=0.1)} & 1.28 \\
9 & Is the legato even? & 52 & \textbf{1.19} & \texttt{SVR(C=0.1)} & 1.21 & \texttt{SVR(C=0.1)} & 1.19 \\
10 & Are the left and right hands balanced? & 31 & \textbf{1.09} & \texttt{RF(n=200)} & 1.36 & \texttt{SVR(C=0.1)} & 1.35 \\
11 & Are the timing onsets aligned on LH and RH? & 29 & \textbf{1.38} & \texttt{SVR(C=10)} & 1.45 & \texttt{RF(n=100)} & 1.57 \\
\bottomrule
\end{tabular}
\caption{Best prediction results from symbolic and audio features. Dummy-MAE is the random guessing baseline.}
\label{tab:algorithm_evaluation}
\end{table*}

\vspace{-5pt}


The dataset includes 2255 audio-question-answer (AQA) triplets, with 391 triplets labeled by one, 874 triplets annotated by two (437 unique), and 990 triplets annotated by three (330 unique) annotators. For those with two and more annotators, we checked for the range between multiple ratings to examine consistency of human judges with this type of question and assessment.For most of the questions, around $65\%$ of the data reached a rough agreement between annotators (with $\leq2$ differences in ratings), although questions on dynamics and legato were the most controversial, with only $50\%$ of the data reaching agreement, and almost no data perfectly agreeing between annotators. Figure~\ref{fig:dist} (top) shows an interpolated CDF plot of the consistency ratings across different questions. The examples that deviates significantly ($\geq5$) are shown in the demo page.

The distribution of teachers' ratings were not uniform, although most of the students were proficient in playing the exercises. The KDE plot in Figure~\ref{fig:dist} bottom shows that most of the questions have opinion peaks in ratings of 2 and 5, indicating a well-separated spread between good and bad performances according to teachers' opinions.

\vspace{-5pt}

\section{Audio and rating}

\begin{figure}
    \centering
    \includegraphics[width=\linewidth]{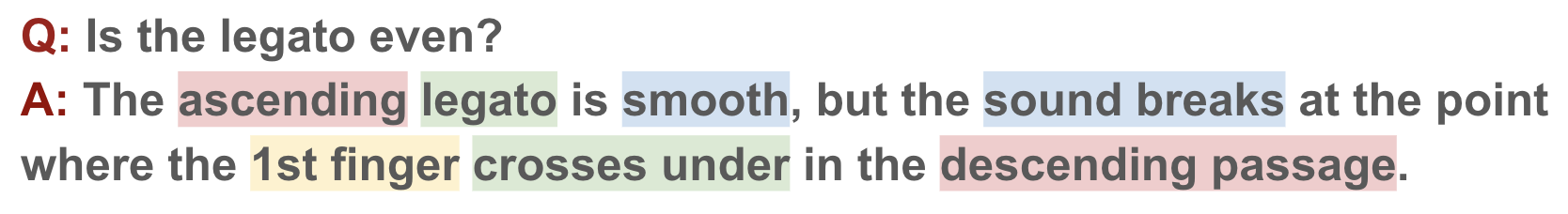}
    \caption{An example of concept annotation in teachers' responses. Red: Location; Yellow: Physicality; Green: Technique; Blue: Description. }
    \label{fig:annot_example}
\end{figure}

\vspace{-5pt}

We explore the relationship between audio performance and rating by predicting the rating with various features using regression. To obtain a convincing training target, we first filtered out responses of the same recording whose ratings differed $>3$ across raters (since that implies a contrast in opinion), and then averaged the remaining ratings
. This left us with 844 AQR (audio-question-rating) triplets. 

Two types of features were extracted: First, audio features were extracted directly from librosa and madmom, namely MFCC-13, chroma, onset envelope, tempogram, and beat estimation. Second, symbolic features, were obtained by first transcribing the clips \cite{Kong2021High-ResolutionTimes}, and then calculating a set of designed features based on partitura \cite{Cancino-Chacon2022PartituraProcessing} note arrays. Then after grouping the notes on the same onset, we computed the inter- and intra-group onset differences, velocity differences and duration differences, aggregated into mean and standard deviation.  

Scipy-implemented regression models, \texttt{SVR}, \texttt{MLP}, \texttt{RandomForest} (\texttt{RF}), and \texttt{GradientBoosting} (\texttt{GB}) were used, with robust scaling, grid search and cross validation as part of the processing pipeline. We have also experimented with different feature selection and combinations, where the best results are presented in Table~\ref{tab:algorithm_evaluation}. For experiment details please refer to our codebase\footnote{https://github.com/anusfoil/neuropiano-data-processing}. The baseline we are comparing against is a dummy regressor that always predicts the mean of the training set.  

Our results show that under traditional MIR feature engineering, not all questions could be fitted. Q8(\textit{dynamics change}) and Q9(\textit{legato}) did not show any improvement from the dummy baseline, but at the same time they were also the most controversial as shown in Section~\ref{sec:introduction}. Likewise, Q2(\textit{solidity}) and Q7(\textit{tonality}) only showed a marginal improvement from dummy baseline.

\begin{figure}
    \centering
    \includegraphics[width=\linewidth]{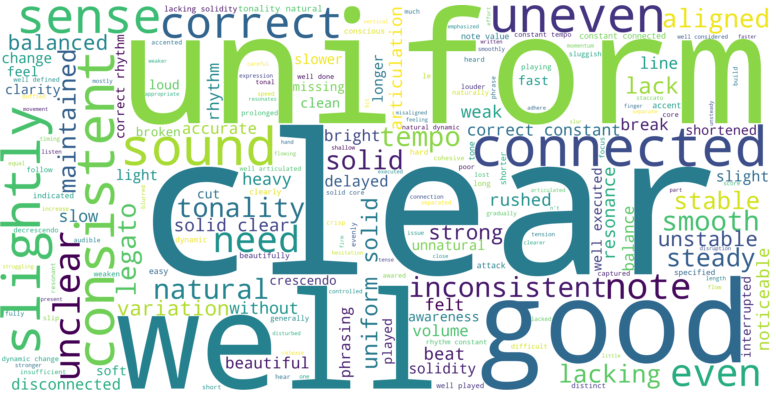}
    \caption{Wordcloud for the descriptions in the dataset.}
    \label{fig:wc_adj}
\end{figure}

\vspace{-5pt}

\section{Ratings and Textual Responses}

We next verified the consistency of teachers' ratings and their implied sentiment, by computing a weighted mean of a 5-point BERT sentiment analysis model as in \cite{Zhang2024LLaQoAssessment}. This yielded a Pearson's correlation of 0.66, for which a box plot of the distribution is shown in our project page\footnote{https://river-blackberry-7de.notion.site/NeuroPiano-data-e59586a44d834ca6bd728a5e2b633880?pvs=4}.


One of the key applications of our dataset is to reveal how teachers comment on performance. Utilizing \texttt{GPT-4o}, we annotated important concepts that were informative within a feedback, namely location, physicality, techniques, and description, as shown in Figure~\ref{fig:annot_example}. 
Observations on 'physicality' gave descriptors such as \textit{left hand}, \textit{thumb}, etc, while the 'location' usually involved \textit{descent}, \textit{beginning}, etc.. Figure~\ref{fig:wc_adj} shows the wordcloud for the prominent descriptions. Future research should leverage this response annotation to gain deeper insights into the feedback process, and we wish the rating, audio and text in NeuroPiano dataset could enhance automated analysis of performance critique. 


\bibliography{ref}

\begin{thebibliography}{10}
\providecommand{\url}[1]{#1}
\csname url@samestyle\endcsname
\providecommand{\newblock}{\relax}
\providecommand{\bibinfo}[2]{#2}
\providecommand{\BIBentrySTDinterwordspacing}{\spaceskip=0pt\relax}
\providecommand{\BIBentryALTinterwordstretchfactor}{4}
\providecommand{\BIBentryALTinterwordspacing}{\spaceskip=\fontdimen2\font plus
\BIBentryALTinterwordstretchfactor\fontdimen3\font minus \fontdimen4\font\relax}
\providecommand{\BIBforeignlanguage}[2]{{%
\expandafter\ifx\csname l@#1\endcsname\relax
\typeout{** WARNING: IEEEtran.bst: No hyphenation pattern has been}%
\typeout{** loaded for the language `#1'. Using the pattern for}%
\typeout{** the default language instead.}%
\else
\language=\csname l@#1\endcsname
\fi
#2}}
\providecommand{\BIBdecl}{\relax}
\BIBdecl

\bibitem{Zhang2024FromPiano}
H.~Zhang, J.~Liang, and S.~Dixon, ``From audio encoders to piano judges: Benchmarking performance understanding for solo piano,'' in \emph{Proceeding of the 25t International Society on Music Information Retrieval (ISMIR)}, 2024.

\bibitem{Kim2022OverviewContext}
H.~Kim, P.~Ramoneda, M.~Miron, and X.~Serra, ``An overview of automatic piano performance assessment within the music education context,'' \emph{International Conference on Computer Supported Education, CSEDU - Proceedings}, vol.~1, 2022.

\bibitem{Eremenko2020PerformanceLearning}
V.~Eremenko, A.~Morsi, J.~Narang, and X.~Serra, ``Performance assessment technologies for the support of musical instrument learning,'' \emph{Proceedings of the 12th International Conference on Computer Supported Education (CSME)}, 2020.

\bibitem{Morsi2023SoundsPerformances}
A.~Morsi, K.~Tatsumi, A.~Maezawa, T.~Fujishima, and X.~Serra, ``Sounds {Out} of {Pl{\"{a}}ce}? {S}core-independent detection of conspicuous mistakes in piano performances,'' in \emph{Proceeding of the 24th International Society on Music Information Retrieval (ISMIR)}, 2023.

\bibitem{Zhang2024LLaQoAssessment}
H.~Zhang, V.~Cheung, H.~Nishioka, S.~Dixon, and S.~Furuya, ``{LLaQo}: Towards a query-based coach in expressive music performance assessment,'' in \emph{Arxiv preprint arXiv:2409.08795}, 2024.

\bibitem{Morsi2024SimulatingLearning}
A.~Morsi, H.~Zhang, A.~Maezawa, S.~Dixon, and X.~Serra, ``Simulating piano performance mistakes for music learning,'' \emph{In Proceedings of the Sound and Music Computing Conference (SMC)}, 2024.

\bibitem{Matsubara2021CROCUSCritiques}
M.~Matsubara, R.~Kagawa, T.~Hirano, and I.~Tsuji, ``{CROCUS}: Dataset of musical performance critiques,'' in \emph{In Proceedings of the International Symposium on Computer Music Multidisciplinary Research (CMMR)}, 2021.

\bibitem{Zhang2024DExterModels}
H.~Zhang, S.~Chowdhury, C.~E. Cancino-Chac{\'{o}}n, J.~Liang, S.~Dixon, and G.~Widmer, ``Dexter: Learning and controlling performance expression with diffusion models,'' \emph{Applied Sciences}, vol.~14, no.~15, 2024.

\bibitem{Lerch2020AnAnalysis}
A.~Lerch, C.~Arthur, A.~Pati, and S.~Gururani, ``An interdisciplinary review of music performance analysis,'' \emph{Transactions of the International Society for Music Information Retrieval}, vol.~3, no.~1, pp. 221--245, 2020.

\bibitem{Zhang2023DisentanglingPerformance}
H.~Zhang and S.~Dixon, ``Disentangling the horowitz factor: Learning content and style from expressive piano performance,'' in \emph{ICASSP 2023 - 2023 IEEE International Conference on Acoustics, Speech and Signal Processing (ICASSP)}, Rhodes Island, Greece, 2023.

\bibitem{Cancino-Chacon2018ComputationalReview}
C.~E. Cancino-Chac{\'{o}}n, M.~Grachten, W.~Goebl, and G.~Widmer, ``Computational models of expressive music performance: A comprehensive and critical review,'' \emph{Frontiers in Digital Humanities}, vol.~5, no. October, pp. 1--23, 2018.

\bibitem{Zhang2022ATEPPPerformance}
H.~Zhang, J.~Tang, S.~Rafee, S.~Dixon, and G.~Fazekas, ``{ATEPP}: A dataset of automatically transcribed expressive piano performance,'' in \emph{Proceedings of the International Society for Music Information Retrieval Conference (ISMIR)}, Bengaluru, India, 2022.

\bibitem{Huang2020Score-informedAssessment}
J.~Huang, Y.-N. Hung, A.~Pati, S.~K. Gururani, and A.~Lerch, ``Score-informed networks for music performance assessment,'' in \emph{Proceedings of the 21st International Society for Music Information Retrieval Conference (ISMIR)}, 2020.

\bibitem{Pati2018AssessmentNetworks}
K.~A. Pati, S.~Gururani, and A.~Lerch, ``Assessment of student music performances using deep neural networks,'' \emph{Applied Sciences (Switzerland)}, vol.~8, no.~4, 2018.

\bibitem{Zhang2021LearnAssessment}
H.~Zhang, Y.~Jiang, T.~Jiang, and P.~Hu, ``Learn by referencing: Towards deep metric learning for singing assessment,'' in \emph{Proceedings of the 22nd International Society for Music Information Retrieval Conference (ISMIR)}, 2021.

\bibitem{Jin2023OrderPerformance}
X.~Jin, W.~Zhou, J.~Wang, D.~Xu, Y.~Rong, and S.~Cui, ``An order-complexity model for aesthetic quality assessment of homophony music performance,'' 2023.

\bibitem{Parmar2021PianoAssessment}
P.~Parmar, J.~Reddy, and B.~Morris, ``Piano skills assessment,'' in \emph{IEEE 23th International Workshop on Multimedia Signal Processing (MMSP)}, 2021.

\bibitem{Jiang2023ExpertFeedback}
Y.~Jiang, ``Expert and novice evaluations of piano performances : Criteria for computer-aided feedback,'' in \emph{Proceeding of the 24th International Society on Music Information Retrieval (ISMIR)}, 2023.

\bibitem{Kong2021High-ResolutionTimes}
Q.~Kong, B.~Li, X.~Song, Y.~Wan, and Y.~Wang, ``High-resolution piano transcription with pedals by regressing onset and offset times,'' \emph{IEEE/ACM Transactions on Audio Speech and Language Processing}, vol.~29, pp. 3707--3717, 2021.

\bibitem{Cancino-Chacon2022PartituraProcessing}
C.~Cancino-Chac{\'{o}}n, S.~D. Peter, E.~Karystinaios, F.~Foscarin, M.~Grachten, and G.~Widmer, ``Partitura: A python package for symbolic music processing,'' in \emph{Proceedings of the Music Encoding Conference (MEC)}, Halifax, Canada, 2022.

\end{thebibliography}





\end{document}